\documentstyle[preprint,psfig,aps]{revtex}
\draft
\tighten
\renewcommand{\vec}[1]{\mbox{\boldmath$#1$\unboldmath}}
\begin{document}

\title{Time-scale of breakup of the halo nucleus $^{11}$Be}

\author{P. Banerjee}
\address{Theory Group, Saha Institute of Nuclear Physics,
\\1/AF Bidhan Nagar, Calcutta - 700 064, INDIA}
\date{\today}
\maketitle

\begin{abstract}
We investigate the post-acceleration of the $^{10}$Be fragment in Coulomb 
breakup reaction of the halo nucleus $^{11}$Be, assuming that the excitation 
of the projectile is to states in the low energy continuum. The method used 
retains all finite-range effects associated with the interactions between the 
breakup fragments and can use realistic wave functions for the halo nuclei. We 
apply the method to compute the average momenta of the $^{10}$Be fragment at
four different scattering angles following breakup of $^{11}$Be on a Au target
at 42 MeV/nucleon beam energy. Our results, which are in qualitative agreement
with recent precision experimental data, are consistent with the picture of
no post-acceleration effect. The implication is that the breakup takes place 
on a time-scale long compared to the collision time.
\end{abstract}


In the breakup reaction of a halo nucleus, there could be a difference in the 
velocity of the charged breakup fragment in the final channel and the beam 
velocity. The magnitude of this velocity difference can be used as a probe to measure the time-scale of such 
reactions \cite{1}. If the breakup takes place at or around the distance of
closest approach, the charged fragment, which is lighter than the projectile, 
is accelerated more on the way out than
the projectile was decelerated on the way in. This phenomenon is often referred
to as ``post-acceleration". On the other hand, if the breakup occurs long after
the fragment has left the strong Coulomb field near the target, the deceleration
and acceleration would be the same. Then the fragment velocity would be equal to
the beam velocity.

The subject of post-acceleration of the heavy charged core in the breakup 
of halo nuclei on heavy targets has been a controversial issue \cite{20}.
Recent measurements on Coulomb dissociation of $^{11}$Li and $^{11}$Be on heavy
targets reveal post-acceleration of the charged cores \cite{2,3,4}. In the measurement of
breakup of $^{11}$Li on Pb target at MSU, a clear non-zero velocity difference
between $^9$Li and the two neutrons has been seen \cite{2}. However, this 
difference has not been observed clearly when the velocities of the $^{10}$Be 
fragment and the
neutron in the final channel have been measured by Nakamura {\em et al.} at 
RIKEN, Japan in the breakup of $^{11}$Be on Pb at 72 MeV/nucleon \cite{3}. But 
significant post-acceleration has been reported by them when the mean 
longitudinal momentum of $^{10}$Be has been measured as a function of its 
scattering angle. Measurements done at MSU on the average parallel momentum
of $^9$Li (resulting from breakup of $^{11}$Li on U) as a function of its 
scattering angle have similar conclusions
regarding post-acceleration of this breakup fragment \cite{4}.  Anne {\em et al.} 
have found little post-acceleration in their measurements of the neutron 
momentum distribution from the fragmentation of $^{11}$Be at GANIL \cite{7}. It has been
argued by them that the post-acceleration effects in these
reactions should be small because the collision time is much less than the
characteristic time for the disintegration of the halo. 

Theoretically, there have been numerous attempts to calculate the post-acceleration
of charged fragments in breakup reactions induced by halo nuclei. The 
observations of Nakamura {\em et al.} agree with the classical calculation 
of Baur, Bertulani and Kalassa \cite{5}, where the concept of a `classical breakup
radius' has been introduced. But the result of this calculation does not agree 
with the angular
dependence of the $^9$Li momentum centroid as observed in the MSU experiment \cite{4}.
This calculation is also in contradiction with the time-dependent, three dimensional
Schr\"odinger calculation of Kido, Yabana and Suzuki \cite{6}. The three dimensional
quantum mechanical calculations of these higher order Coulomb final state effects
by Esbensen, Bertsch and Bertulani \cite{1} do not support the argument of Anne
{\em et al.} \cite{7}.  However, post-acceleration
effects noticed in the energy distributions of charged breakup fragments
calculated at different angles within the theory of post form DWBA Coulomb 
breakup are insignificant \cite{20,8}. There the authors are of opinion that the breakup takes place at
large distances because of the small separation energy of the valence neutron(s).
Thereby the effect of repulsion of the strong Coulomb field is reduced, producing
no appreciable post-acceleration \cite{20,8}. No post-acceleration has also been
observed in the parallel momentum distribution of $^{10}$Be resulting from 
calculations using an adiabatic model of Coulomb breakup of $^{11}$Be 
on heavy (U and Ta) targets \cite{9}. Thus, different observables have been
calculated in this regard with different conclusions and theoretically also
the observation of post-acceleration remains debatable.

In order to verify the experimental result of Nakamura {\em et al.}, precise 
measurements have been done very recently at MSU \cite{18}. In this brief report, we calculate the 
post-acceleration of the charged fragment $^{10}$Be in the Coulomb dissociation
of the halo nucleus $^{11}Be$ and compare with the MSU data. We follow the 
theoretical formalism which was first described in \cite{10} for the Coulomb 
breakup of a light weakly bound two-body composite nucleus $a$ consisting of a charged core $c$ and a 
neutral valence particle $v$ on target $t$ at energies of a few tens of MeV per
nucleon and above. 
There are two approximations used in this theory - that the dominant projectile
breakup configurations excited are in the low-energy continuum and that the
valence particle does not interact with the target. The theory is fully
quantum mechanical and is also non-perturbative. The method retains all 
finite-range effects associated with the interactions among the breakup 
fragments and includes the initial and final state interactions to all orders. 
It allows the use of realistic wave functions to describe the halo nuclei.

The transition amplitude for the above elastic Coulomb breakup reaction, in the
c.m. frame, is  given by (Fig. 1)
\begin{equation}
T_{\sigma _c\sigma _v;\sigma _a}=\langle \chi^{(-)} (\vec{k}_c, \vec{R}_c)
{\cal S}_{\sigma _c} e^{i\vec{k}_v\cdot\vec{R} _v}{\cal S}_{\sigma _v}\vert
V_{cv}\vert \Psi ^{(+)}_{\vec{k}_a{\sigma _a}} (\vec{r},\vec{R})\rangle~,
\label{tmat}
\end{equation}
where ${\cal S}_{\sigma _c}$ and ${\cal S}_{\sigma _v}$ are the core and
valence particle internal wavefunctions with $\sigma_c$ and $\sigma_v$ their
spin projections. $\hbar\vec{k}_c$ and $\hbar\vec{k}_v$ are the asymptotic
momenta of these fragments, conjugate to $\vec{R}_c$ and $\vec{R}_v$,
respectively, and $\chi^{(-) }$ is an in-going waves Coulomb distorted wave
function describing the $c$--$t$ relative motion in the final state. Since it
is assumed that $V_{vt}=0$ the valence particle is described by a plane wave in
the final state.

Following the adiabatic approximation of ref.\ \cite{11}, the exact three-body
scattering wave function $\Psi ^{(+)}_{\vec{k}_a{\sigma _a}} (\vec{r},\vec{R})$
separates in the variables $\vec{R}_c$ and $\vec{r}$, namely
\begin{equation}
\Psi ^{(+)}_{\vec{k}_a\sigma _a}
(\vec{r},\vec{R}) \approx \Psi_{\vec{k}_a{\sigma _a}}^{\rm AD}(\vec{r},\vec{R})
= \Phi _{a\sigma _a}(\vec{r}) e^{i\gamma\vec{k}_a\cdot \vec{r}} \chi
^{(+)}(\vec{k}_a,\vec{R}_c)~.
\label{adwf}
\end{equation}
Here $\chi ^{(+)}$ is a Coulomb distorted wave representing projectile's
motion in the incident channel, evaluated at the core coordinate $\vec{R}_c$
and $\gamma = m_v/(m_c + m_v)$.  

The projectile ground state wave function is given by 
\begin{eqnarray}
\Phi _{a\sigma _a}(\vec{r})=\sum_{l\mu jm\sigma_c'\sigma_v'} \langle s_c \sigma
_c'jm \vert s_a\sigma _a\rangle\langle l\mu s_v\sigma _v'\vert jm \rangle
\Phi_{a}^{l\mu}(\vec{r}){\cal S}_{\sigma _c'}{\cal S}_{\sigma _v'}~,
\end{eqnarray}
where $\Phi _{a}^{l\mu}(\vec{r})= i^l u_l(r) Y_{l\mu}(\hat{\vec{r}})$, the
$u_l$ are radial wavefunctions, and the $Y_{l\mu}$ are the spherical
harmonics.  Since the only distorting interaction $V_{ct}$ is assumed to be central,
the integrations over spin variables can be carried out in Eq.\ (\ref{tmat}).
The required approximate transition amplitude can then be expressed as
\begin{eqnarray}
T^{\rm AD}_{\sigma _c\sigma _v;\sigma _a}=\sum_{l\mu j m} \langle s_c\sigma _cj
m \vert s_a\sigma _a\rangle\langle l\mu s_v\sigma _v\vert jm \rangle 
\beta^{AD}_{l\mu}~,
\end{eqnarray}
where the reduced transition amplitude $\beta^{AD}_{l\mu}$ is 
\begin{eqnarray}
\beta^{\rm AD}_{l\mu}=\langle\chi ^{(-)}(\vec{k}_c,
\vec{R}_c) e^{i\vec{k}_v \cdot \vec{R}_v} \vert V_{cv}\vert \Phi
_{a}^{l\mu}(\vec{r}) e^{i\gamma \vec{k}_a \cdot \vec{r}} \chi
^{(+)}(\vec{k}_a,\vec{R}_c) \rangle~.
\end{eqnarray}
Since $\vec{R}_v= \alpha\vec{R}_c + \vec{r}$ (Fig. 1), where $\alpha = m_t/(m_t
+m_c)$, then without further approximation the entire adiabatic amplitude now
separates exactly in the coordinates $\vec{R}_c$ and $\vec{r}$, as
\begin{eqnarray}
\beta^{\rm AD}_{l\mu}&=&\langle e^{i\vec{q}_v\cdot \vec{r}}\vert V_{cv}\vert
\Phi _{a}^{l\mu}(\vec{r}) \rangle \langle \chi ^{(-)}(\vec{k}_c,
\vec{R}_c)e^{i\alpha\vec{k}_v\cdot \vec{R}_c} \vert \chi
^{(+)}(\vec{k}_a,\vec{R}_c) \rangle~\nonumber \\ 
&=&\langle \vec{q}_v\vert V_{cv}\vert \Phi _{a}^{l\mu}\rangle
\langle \chi ^{(-)}(\vec{k}_c);\alpha\vec{k}_v\vert \chi
^{(+)}(\vec{k}_a) \rangle~.\label{betad}
\end{eqnarray}
The momentum $\vec{q}_v$ appearing in the first term is $\vec{q}_v
=\vec{k}_v-\gamma\vec{k}_a$.

Here the structure information about the projectile
is contained only in the first term, the vertex function, denoted by 
$D(\vec{q}_v) = D_l(q_v)Y_{l\mu}(\hat{\vec{q}}_v)$, where
\begin{eqnarray}
D_l(q)= 4\pi\int^{\infty}_0 dr r^2 j_l(qr)V_{cv}(r) u_l(r)~.
\end{eqnarray}
The second factor is associated with the dynamics of the reaction only, which
is expressable in terms of the bremsstrahlung integral \cite{12}.

The triple differential cross section for the elastic breakup reaction is
\begin{eqnarray}
{d^3\sigma \over dE_c d\Omega _cd\Omega _v}={2\pi\over \hbar v_a}\left\{{1\over
2s_a + 1}\sum_{\sigma _c\sigma _v\sigma _a}\vert T^{\rm AD}_{\sigma _c\sigma _v;
\sigma _a}\vert ^2\right\} \rho (E_c,\Omega _c,\Omega _v)  ~, \label{trip}
\end{eqnarray}
or, upon carrying out the spin projection summations,  
\begin{eqnarray}
{d^3\sigma \over dE_c d\Omega _cd\Omega _v}={2\pi\over \hbar v_a}\left\{\sum_{l
\mu}\frac{1}{(2l + 1)}\vert \beta^{\rm AD}_{l\mu}\vert^2 \right\} \rho(E_c,
\Omega _c,\Omega _v)~.
\end{eqnarray}
Here $v_a$ is the $a$--$t$ relative velocity in the entrance channel.
The phase space factor $\rho (E_c,\Omega _c,\Omega _v)$ appropriate to the
three-body final state is \cite{13,14}
\begin{eqnarray}
\rho (E_c,\Omega _c,\Omega _v) = {h^{-6}m_tm_cm_vp_cp_v\over m_v + m_t
-m_v\vec{p}_v\cdot(\vec{P} -\vec{p}_c)/p_v^2} \label{phas}
\end{eqnarray}
where, for the differential cross section in the laboratory frame, $\vec{P}$,
$\vec{p}_c$ and $\vec{p}_v$ are the total, core, and valence particle momenta in
the laboratory system.

The core three
dimensional momentum distribution $d^2\sigma \over dp_c d\Omega _c$ is related
to its energy distribution cross section by
\begin{eqnarray}
{d^2\sigma \over dp_c d\Omega _c} = \sqrt{2E_c \over m_c}{d^2\sigma \over dE_c 
d\Omega _c}
\end{eqnarray}
${d^2\sigma \over dE_c d\Omega _c}$ can be readily obtained from the triple
differential cross section above by integration with respect to the solid
angle of the valence particle.

Both theoretically \cite{15,16} 
and experimentally \cite{17}, $^{11}$Be is known to have a dominant 1$s_{1\over 
2}$ neutron configuration in its ground state. We have calculated the breakup 
of $^{11}$Be assuming a 1$s_{1\over 2}$ neutron orbital with separation energy
0.504 MeV. The binding potential for $^{11}$Be is assumed to be of 
Woods-Saxon type with radius and diffuseness parameters 1.15 fm and 0.5 fm
respectively.  The depth of this potential has been 
calculated to reproduce the binding energy. 

Very recently, the three dimensional momentum distribution
$d^2\sigma \over dp_cd\Omega _c$ has been measured as a function of the 
momentum $p_c$ of the $^{10}$Be core at MSU from elastic breakup of $^{11}$Be 
around 42 MeV/nucleon (41.71 MeV/nucleon) beam energy on a Au target at four 
different angles of 0.64$^{\circ}$, 1.50$^{\circ}$, 2.88$^{\circ}$ and 4.39$^
{\circ}$ \cite{18,21}. The fact that the target remains in the ground state has 
been ensured by observing no $\gamma$-rays in coincidence, which might result 
as a consequence of the de-excitation of the possibly excited target. The 
average momentum has been computed in each case. 

We have calculated the 
momentum distributions of the $^{10}$Be fragment in the Coulomb breakup of 
$^{11}$Be on Au at 42 MeV/nucleon incident energy at these four angles. These
have been shown in Fig. 2. The breakup of $^{11}$Be on high $Z$ targets is known
to be Coulomb dominated and this is even more true at forward angles 
\cite{20,7,8,9}. Therefore, we have not considered nuclear contributions in our calculations.
The average momentum at each angle has been calculated
by using the expression $\sum p_c{d^2\sigma \over dp_cd\Omega _c}/\sum
{d^2\sigma \over dp_cd\Omega _c}$. The post-acceleration should show up in the 
increase of this average momentum with increase of scattering angle. This is 
because with increase of scattering angle, the classical impact parameter 
decreases and the Coulomb repulsion on the outgoing charged fragment increases. 
In Fig. 3, we show results of our calculated average momenta along with 
calculations done for $^{11}$Be at 42 MeV/nucleon beam energy using 
methods of classical theory presented in \cite{5}. With 42 MeV/nucleon 
incident energy, the beam velocity momentum is approximately 2806 MeV/c. Our 
calculated average momenta come around the same value. Therefore, we do not 
see any post-acceleration. The classical calculations, on the other hand, favour 
post-acceleration (Fig. 3). In our calculations, the Coulomb interactions in the 
final channel are included to all orders. It should be mentioned that higher 
order semi-classical theory in \cite{19} and quantal theory with high energy 
approximation in \cite{5} also did not support post-acceleration of $^9$Li in 
the breakup of $^{11}$Li. The experimental data, consistent with a constant
average momentum over the range of scattering angles measured, also do not 
support the observation of this phenomenon. However, the experimentally deduced 
mean momenta (Fig. 3) are somewhat larger than 2797 MeV/c, which is the beam 
velocity momentum corresponding to the experimental beam energy of 41.71 
MeV/nucleon. 

In conclusion, our calculations on average momentum of the $^{10}$Be fragment
at four different scattering angles following the Coulomb breakup of the
one-neutron halo nucleus $^{11}$Be on a Au target at 42 MeV/nucleon beam energy
gives no evidence of post-acceleration of $^{10}$Be. The calculations are 
performed within an approximate quantum mechanical theoretical model of elastic
Coulomb breakup, which makes
the following assumptions: (i) only the charged core interacts with the target
via a point Coulomb interaction and (ii) that the important excitations of the
projectile are to the low-energy continuum, and so can be treated adiabatically.
The method permits a fully finite-range treatment of the projectile vertex and
includes initial and final state interactions to all orders. The calculated 
results are compatible with the findings of recent good quality measurements
at MSU. The results are consistent with a picture in which 
the breakup of $^{11}$Be takes place on a time-scale long compared to the 
collision time.

\acknowledgments 
The author would like to thank Dr. J. A. Tostevin and Dr. I. J. Thompson of
the University of Surrey, U.K. for stimulating 
discussions. Thanks are also due to Prof. R. Shyam for kindly going through 
this manuscript and making valuable suggestions.

\begin{figure}
\centerline{\psfig{file=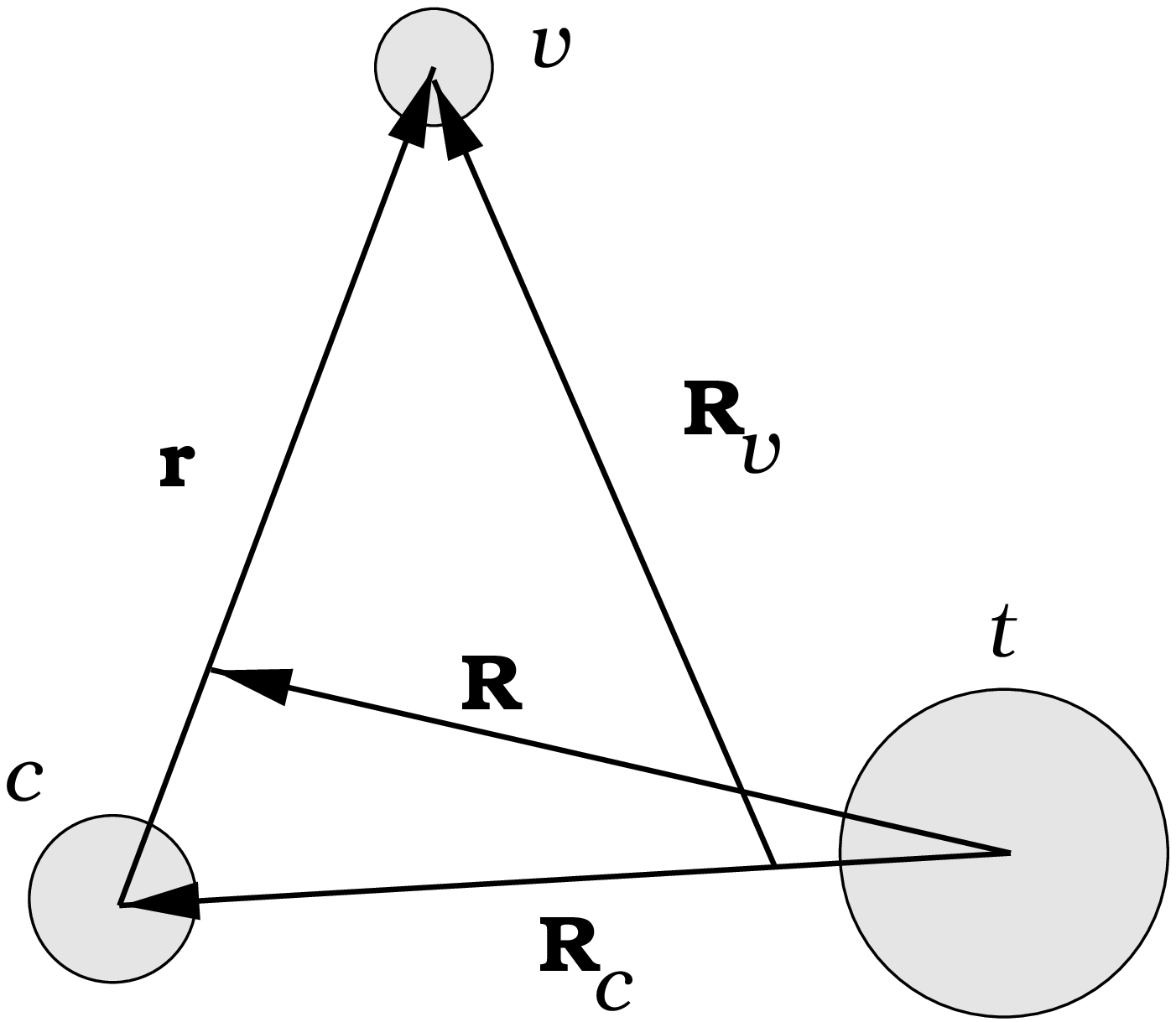,width=0.7\textwidth}}
\caption{Coordinate system adopted for the core, valence particle and
target three-body system.}
\end{figure}

\begin{figure}
\centerline{\psfig{file=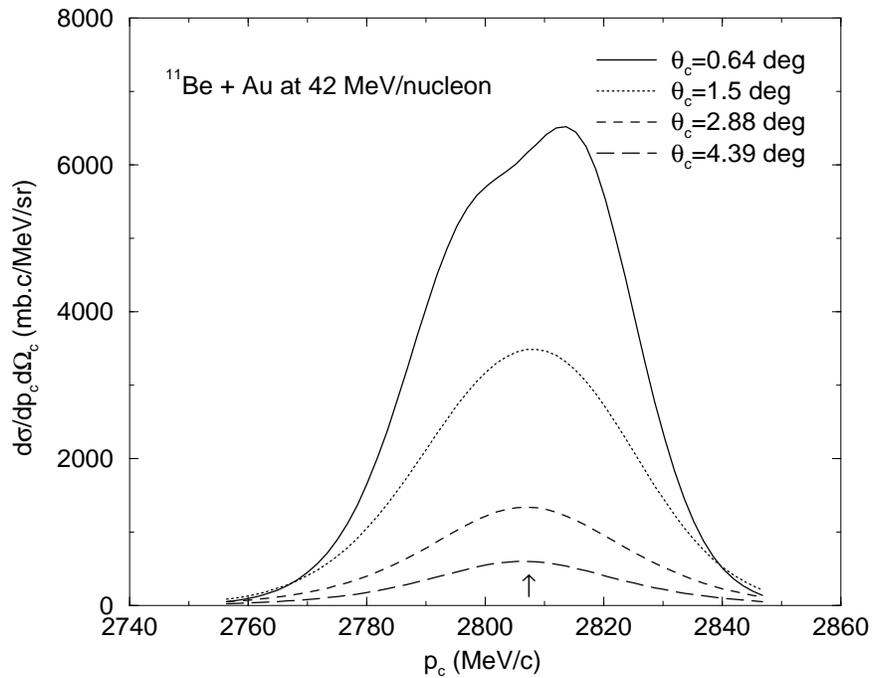,width=0.7\textwidth}}
\caption{Calculated three dimensional momentum distribution of $^{10}$Be from Coulomb breakup 
of $^{11}$Be on Au at 42 MeV/nucleon at four different laboratory angles
of scattering of $^{10}$Be. 
The arrow on the horizontal axis indicates the position of the beam velocity
momentum.}
\end{figure}

\begin{figure}
\centerline{\psfig{file=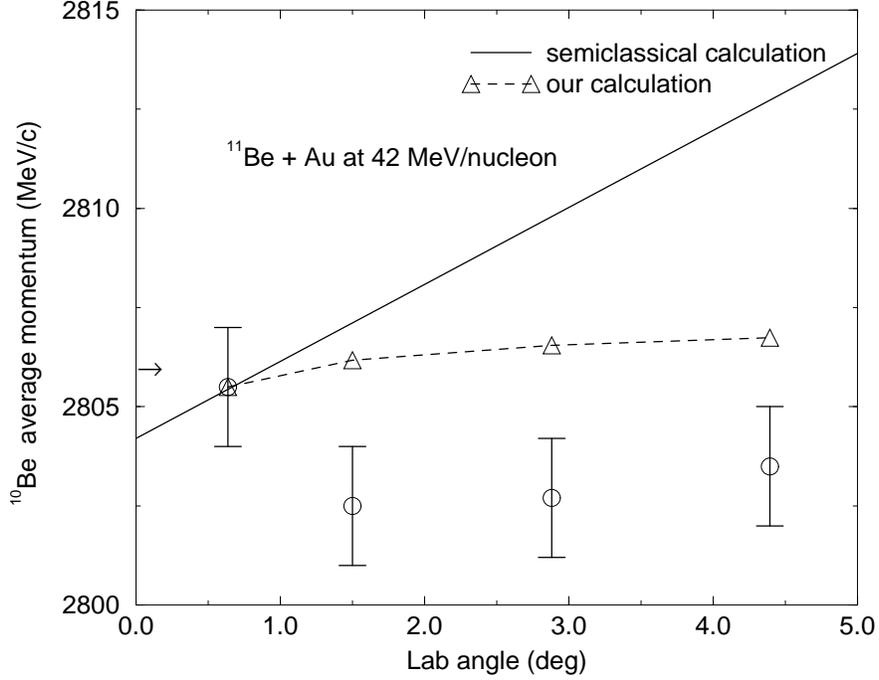,width=0.7\textwidth}}
\caption{Calculated average momentum of $^{10}$Be from Coulomb breakup of
$^{11}$Be on Au at 42 MeV/nucleon as a function of the laboratory angle
of scattering of $^{10}$Be. The experimental data, taken from \protect\cite{18},
are at 41.71 MeV/nucleon. The semiclassical calculations are results of 
formulae in \protect\cite{5}.
The arrow on the vertical axis indicates the position of the beam velocity
momentum.}
\end{figure}
\end{document}